# Electrical detection of antiferromagnetic dynamics in Gd-Co thin films by using a 154-GHz gyrotron irradiation

S. Funada[1], Y. Ishikawa[2], M. Kimata[3], K. Hayashi[2], T. Sano[2], K. Sugi[1], Y. Fujii[2], S. Mitsudo[2], Y. Shiota[1,4], T. Ono[1,4], and T. Moriyama*[1,4,5]

[1] *Institute for Chemical Research, Kyoto Univ., Uji, Kyoto 611-0011, Japan*
[2] *Research Center for Development of Far-Infrared Region, Univ. of Fukui, Fukui, Fukui 910-8507, Japan*
[3] *Institute for Materials Research, Tohoku Univ., Sendai, Miyagi 980-8577, Japan*
[4] *Center for Spintronics Research Network, Kyoto Univ., Uji, Kyoto 611-0011, Japan*
[5] *PRESTO, Japan Science and Technology Agency, Kawaguchi, Saitama 322-0012, Japan*

**Abstract**

THz magnetization dynamics is a key property of antiferromagnets as well as ferrimagnets that could harness the THz forefront and spintronics. While most of the present THz measurement techniques are for bulk materials whose sensitivities rely on the volume of the material, measurement techniques suitable for thin films are quite limited. In this study, we explored and demonstrated electrical detection of the antiferromagnetic dynamics in ferrimagnetic Gd-Co thin films by using a 154 GHz gyrotron, a high-power electromagnetic wave source. Captured resonant modes allow us to characterize the peculiar magnetization dynamics of the Gd-Co around the net angular momentum compensation. As the gyrotron frequency is scalable up to THz, our demonstration can be an important milestone toward the THz measurements for antiferro- and ferri- magnetic thin films.

*Corresponding to *mtaka@scl.kyoto-u.ac.jp*

Antiferromagnets are one of the promising magnetic materials which can sustain and work at THz frequency[1]. Amid the antiferromagnetic-ferrimagnetic spintronics[2,3,4,5] where antiferromagnets and ferrimagnets are used as active materials in spintronic devices, the THz magnetization dynamics of antiferromagnets and ferrimagnets is a key property that could harness the THz forefront and spintronics. Recent advancements in THz measurement techniques have realized some of the key experiments on the antiferromagnetic dynamics and the relevant phenomena, such as control of THz spin oscillation[6,7,8], THz spin pumping effect[9], and ultrafast spin switching[10,11].

However, the measurement principles used in those studies, such as the magneto-optical effect and resonant absorption, rely on the volume of the materials. Therefore, the same principle can hardly be applied for characterizing the dynamics of the thin films which is a central interest when considering any antiferro- and ferri-magnetic integration devices. Indeed, due to the measurement difficulty, basic properties of magnetization dynamics in antiferromagnetic thin films, such as resonant frequency and magnetic damping constant, remain elusive while it is not perhaps legitimate to assume that those properties are same as those in bulk materials. Therefore, establishment of the measurement technique for characterizing the antiferromagnetic dynamics in thin films is a pressing issue for THz spintronics.

Electrical detection of the magnetic resonance is one of the routes to establishment of such a measurement technique. For ferromagnetic thin films, there are wide varieties of the electrical measurement techniques that capture the magnetization dynamics. For instance, a homodyne detection technique taking advantage of a nonlinear coupling between a rf current and a magnetoresistance oscillation[12,13,14] is widely used to characterize the dynamics in ferromagnetic thin films. The resultant electrical signal can be expressed as $V = (\Delta\rho/\mathbf{M}^2)(\mathbf{J} \cdot \mathbf{M})\mathbf{M} - R_H \mathbf{J} \times \mathbf{M}$, where $\Delta\rho$ is the resistance change due to magnetoresistance, $\mathbf{M}$ is the magnetization, $\mathbf{J}$ is the induction current by the electromagnetic wave irradiation, $R_H$ is the anomalous Hall effect constant. The technique can not only characterize the ferromagnetic films as thin as sub-nano meters[15] but also can characterize various interesting nonequilibrium phenomena such as spin-torque[16]. Another is a spin-to-charge conversion technique taking advantage of a combination of the spin pumping effect and the inverse spin Hall effect in a bilayer of ferromagnet/spin Hall metal, where the spin Hall metal is Pt, Ta, etc[17,18]. Persistent magnetization dynamics in the ferromagnetic layer creates and pumps a spin current into the spin Hall metal layer by the spin pumping effect. The spin current injected in the spin Hall metal layer is converted to a charge current by which the electrical detection of the magnetization dynamics is realized. The principle of the latter technique has recently been demonstrated in particular bulk-antiferromagnet/spin Hall metal interfaces[19,20], i.e., $MnF_2$/Pt and $Cr_2O_3$/Pt, which strongly suggests the characterization of the thin films is possible.

An electrical signal, or more precisely a dc voltage signal, emerging at the resonance for both techniques is in principle proportional to a power of the input electromagnetic wave. Considering the case for ferromagnets, a sizable input microwave power (>mWatt) is generally required in order to obtain a comfortably detectable voltage signal of the order of μV[14,18]. A concern when it comes to the THz frequency is that such high-power continuous wave (CW) THz sources are quite limited. Gyrotron is one of the few CW THz sources which can produce a power of >mWatt[21]. It consists of a linear-beam vacuum tube in which the electron beam undergoes the cyclotron resonance in a strong magnetic field. The cyclotron motion of the electrons emits electromagnetic waves at the resonant frequency as well as its higher harmonics. It is a lab-size equipment and more accessible than other technologies such as free-electron laser[22].

In this work, we explore electrical detection of antiferromagnetic resonant modes in $Gd_xCo_{1-x}$/Ta and $Gd_xCo_{1-x}$/Pt thin films by employing a gyrotron which can generate 154 GHz with a nominal power of 500 mW. We clearly detected the resonances as a voltage signal over which the characteristic magnetization dynamics of the Gd-Co alloy and its temperature dependence are discussed.

Bilayers of $Gd_{0.17}Co_{0.83}$ 20 nm/Ta 3 nm (Sample A), $Gd_{0.16}Co_{0.84}$ 20 nm/Ta 3 nm (Sample B), and $Gd_{0.16}Co_{0.84}$ 20 nm/Pt 3 nm (Sample C) were deposited on a thermally oxidized Si substrates by a magnetron sputtering with a base pressure of $1.5 \times 10^{-6}$ Pa. $Gd_xCo_{1-x}$ layers were formed by co-sputtering of Gd and Co and the composition was controlled by the sputtering power for each element. Ta and Pt capping layer is to protect the Gd-Co layer from oxidation as well as is to utilize the inverse spin Hall effect. Gd-Co alloys are one of the typical rare earth-transition metal ferrimagnets which show both magnetization and angular momentum compensation temperatures due to the difference in the temperature dependences of the magnetic moment and the gyromagnetic ratio of the Gd and Co. The Gd-Co alloys generally show two dynamic modes due to the antiferromagnetically coupled magnetic sublattices of Gd and Co [23,24]. Peculiar magnetization dynamics in the vicinity of those compensation temperatures are comprehended by non-zero angular momentum with zero magnetization and vice versa[25,26,27,28].

Figure 1(a) shows the schematic of the measurement setup with a gyrotron. 154 GHz continuous electromagnetic wave with a nominal power of 500 mW is generated by a gyrotron and is guided into a hollow waveguide sticking out of a cryostat with a superconducting magnet. The electromagnetic wave is linearly polarized by a wire grid placed in a pathway. The sample shaped into a 1.5 mm × 5 mm rectangular piece is placed at the end of the waveguide in the way that the Poynting vector of the electromagnetic wave is parallel both to the sample surface and the sample transverse direction. The linearly polarized rf h-field is aligned perpendicular to the sample plane. Voltage is measured across the sample longitudinal direction with sweeping external magnetic field $H_{ext}$ colinear to the Poynting vector. Field direction opposite to the Poynting vector is defined as the positive field as shown in Fig. 1 (a).

Figure 1(b) shows typical spectra for Sample A with the Ta capping layer at room temperature with the positive and negative magnetic field. Clear dc voltage peaks are observed and there are two resonant modes at around ±2 T and ±4 T. The peaks at around ±4 T (labeled as ⋆) are larger than those at around ±2 T (labeled as Δ) and the polarity of the peak voltage is opposite to each other. For both modes, polarity of the peak voltage is reversed when the magnetic field is reversed, which suggests that the dc voltage is from magnetic resonances of the Gd-Co.

First, we discuss the origin of the dc voltage peak. As discussed above, there are two possible detection mechanisms in our measurement configuration, i.e., the homodyne mechanism and the spin pumping mechanism, that create the dc voltage at the resonance. If the spin pumping mechanism is dominant, Sample B with the Ta capping layer and Sample C with the Pt capping layer would show the opposite voltage polarity at the resonance because sign of the spin Hall angle for Pt and Ta is opposite to each other[29,30]. As shown in Fig. 1 (c), the voltage polarity at the resonance is the same for both Sample B and C, indicating a dominant mechanism is not the spin pumping but the homodyne caused by the nonlinear coupling between the induction current and the magnetoresistance of the Gd-Co film. A slight difference in the line shape between samples could come from a misalignment of the rf h-field with respect to the sample geometry as the line shape is quite sensitive to the alignment with the homodyne mechanism[13,31].

Next, we discuss temperature dependence of the resonant modes. Figure 2 (a) shows the spectra for Sample A at various temperatures. We observed a clear dc voltage peak shifting in the range between 4 and 2 T with decreasing temperature. Figure 2 (b) summarizes the resonant field $\mu_0 H_{res}$ and resonant linewidth $\mu_0 \Delta H_{res}$, which are extracted by the Lorentzian peak fitting, as a function of temperature. $\mu_0 H_{res}$ decreases with decreasing the temperature and $\mu_0 \Delta H_{res}$ varies with respect to the temperature and took its maximum at round 100 K.

Associating with the temperature dependent magnetization as well as the temperature dependent angular momentum, Gd-Co alloys generally show a strong temperature dependence of the resonant frequency. We here define the temperature dependent net magnetization $M(T)$ as well as net angular momentum $s_{net}(T)$ as[32],

$$M(T) = M'_{Co}(1 - T/T_c)^{\beta_{Co}} - M'_{Gd}(1 - T/T_c)^{\beta_{Gd}} \quad (1)$$

$$s_{net}(T) = M'_{Co}/\gamma_{Co}(1 - T/T_c)^{\beta_{Co}} - M'_{Gd}/\gamma_{Gd}(1 - T/T_c)^{\beta_{Gd}} \quad (2)$$

where $T_c$ is the critical temperature for the ferrimagnetic-paramagnetic transition, and $M'_{Gd,Co}$ and $\beta_{Gd,Co}$ are the magnetization at 0 K and the critical exponent, respectively, for each element. $\gamma_{Gd,Co} = g_{Gd,Co}\mu_B/\hbar$ is the gyromagnetic ratio with the Bohr magneton $\mu_B$, the reduced Planck's constant $\hbar$, and the g-factor $g_{Gd,Co}$ for each element. By obtaining the $M'_{Gd,Co}$ and $\beta_{Gd,Co}$ from measured $M(T)$, $s_{net}(T)$ can be estimated using the literature value of $g_{Gd}$= 2.02 and $g_{Co}$= 2.11[33].

Figure 3 shows the measured $M(T)$ for Sample A with the magnetization compensation temperature $T_M \sim 40$ K. The obtained $s_{net}(T)$ is overlayed in Fig. 3. The angular momentum compensation temperature $T_A$ is found to be 130 K. Resonant linewidths in Gd-Co are theoretically predicted to increase as the temperature approaches to $T_A$. As we saw in Fig. 2 (b), the $\mu_0 \Delta H$ peaking at 100 K roughly corresponding to the estimated $T_A$ is consistent with the theory [34] and the previous experimental observations[23,35].

Ferrimagnetic dynamics is essentially described by two sets of Landau-Lifshitz-Gilbert (LLG) equations for each magnetic sublattice of Gd and Co. When solving the simultaneous equation using the Neel vector basis, one obtains the eigenfrequencies[36,37],

$$(f^\pm)^2 = \frac{s_{net}^2 + \rho \mu_0 M(2H_{ext} + M) \pm \sqrt{s_{net}^4 + 2\rho s_{net}^2 \mu_0 M(2H_{ext} + M) + \rho^2(\mu_0 M)^2(M)^2}}{2(2\pi\rho)^2} \quad (3)$$

where $f^-$ and $f^+$ indicate the low frequency mode and the high frequency mode, respectively, and $\rho$ is the momentum of inertia[38]. $\mu_0$ is the vacuum permeability. With Eq. 3, $s_{net}(T)$ can also be obtained at each temperature by using the obtained $H_{res}$, $M(T)$, and $f^\pm = 154$ GHz with a literature value of $\rho = 2 \times 10^{-19}$ kg·m$^{-1}$ [37].

Figure 4 shows $s_{net}(T)$ estimated from Eq. 3 and that estimated from Eq. 2 by the measured $M(T)$ (taken from Fig. 2). Data points of the blue and red squares are obtained for the larger resonant peaks (labeled as ★) by setting $f^- = 154$ GHz and $f^+ = 154$ GHz, respectively. There were no solutions of $s_{net}(T)$ for the smaller peaks (labeled as △), indicating these peaks could be associated with non-uniform magnetization modes, or spin wave modes. $s_{net}(T)$ obtained by two different ways are roughly consistent to each other. In addition, this calculation identifies that observed resonant peaks are from the low frequency mode above $T_A$ and high frequency mode below $T_A$.

These results invoke an immediate question why there is only one mode at given temperatures. In order to address this issue, we introduce a concept of handedness for the magnetization dynamics. When solving the set of LLG equations using the elliptical basis of each sublattice magnetization, one then finds $f^- = f_R$ and $f^+ = f_L$ at $T > T_A$ and $f^- = f_L$ and $f^+ = f_R$ at $T_M < T < T_A$ comparing with Eq. 3[24], where $f_R$ and $f_L$ are, respectively, the resonant frequency for the right-handed and left-handed precession mode with respect to a principal axis parallel to the net magnetization. This analysis indeed reveals the resonant peaks we observed are always the right-handed modes in all the temperature range. We therefore find out that the left-handed modes are invisible in our experiment.

In most of the previous experimental studies on the Gd-Co dynamics, on the contrary to our results, the two modes at a given temperature have been clearly identified [23,24]. We suspect the reason why we do not see the left-handed modes might be rooted in how the dynamics is excited. Those previous studies used either a pump-probe technique or the Brillouin light scattering (BLS). The pump-probe technique thermally excites the magnetization dynamics and the BLS does by the energy and momentum transfer from photons, the both excitation mechanisms of which are not directly relevant to the magnetic susceptibility. On the other hand, in the present measurement, the dynamics is purely driven by the rf h-field. Therefore, magnetic susceptibility is an important factor to efficiently excite the magnetization dynamics in the present study. Indeed, the magnetic susceptibility at $f_L$ is found to be smaller than that at $f_R$ and they can differ by as large as a factor of 10 (see the supplementary material[39,40,41,42,43,44,45]), which could explain the resonant peaks of the left-handed mode are too small to be seen in our experiment.

Finally, we would like to note on the measurement configuration. Since the homodyne mechanism seems to be dominant in our measurement, it is possible to further enhance and maximize the dc voltage signal by looking into a right geometry of the applied field direction and the rf h-field directions with respect to the sample[14,31], which in the present study cannot be realized due to various geometrical limitation in the measurement setup but will be optimized in the future work.

In summary, we demonstrated the electrical detection of magnetic resonance in Gd-Co ferrimagnetic thin films at 154 GHz by employing the gyrotron as a high-power electromagnetic wave source. It was found that the dc voltage peaks are predominantly from the homodyne mechanism. The temperature dependence of the resonant modes is well explained by the temperature dependence of the net angular momentum in the Gd-Co. Lack of the left-handed resonant modes in our observation could be related to the significant difference in the magnetic susceptibility of the right- and left-handed modes. Since the homodyne detection is common effect in magnetic films with appreciable magnetoresistance and the gyrotron frequency is scalable up to THz[21], our demonstration can be an important milestone toward the THz measurements for antiferro- and ferri-magnetic thin films.

**Acknowledgments**

This work was supported by JSPS KAKENHI Grant Numbers 20H05665, 21H04562, 21J10136, and 22H01936, JST PRESTO Grant Number JPMJPR20B9, and the Mazda Foundation. This work is partly supported by the Cooperative Research Program of the Research Center for Development of Far-Infrared Region, University of Fukui (No. R03FIDG036A, R04FIRDM025B).

**Figure Captions**

**Figure 1** (a) Schematic illustration of the measurement setup. 154 GHz electromagnetic wave is generated by the gyrotron and irradiated to the sample. Dc voltage across the sample is measured with sweeping external magnetic field. (b) Typical dc voltage spectra with the positive and negative magnetic field for Sample A ($Gd_{0.17}Co_{0.83}$ 20 nm/Ta 3 nm) at room temperature. (c) Dc voltage spectra for Sample B ($Gd_{0.16}Co_{0.84}$ 20 nm/Ta 3 nm) and Sample C ($Gd_{0.16}Co_{0.84}$ 20 nm/Pt 3 nm) at room temperature. ⋆ and Δ are labeled on the larger and smaller peaks, respectively.

**Figure 2** (a) Dc voltage spectra at various temperatures for Sample A. (b) Temperature dependence of the resonant field $\mu_0 H_{res}$ and linewidth $\mu_0 \Delta H_{res}$ for the resonant peaks labeled as ⋆ and Δ. The data markers correspond to the labels of Fig. 1 (b).

**Figure 3** Measured $M(T)$ for Sample A under the in-plane magnetic field of 100 mT and the estimated $s_{\text{net}}(T)$ by Eq. 2.

**Figure 4** Comparison of $s_{\text{net}}(T)$ estimated by Eq. 2 and Eq. 3. The blue and red squares are obtained for the resonant peaks (labeled as ⋆ in Fig. 2 (a)) by setting $f^- = 154$ GHz and $f^+ = 154$ GHz in Eq. 3, respectively. The black line shows the estimated $s_{\text{net}}(T)$ by Eq. 2.

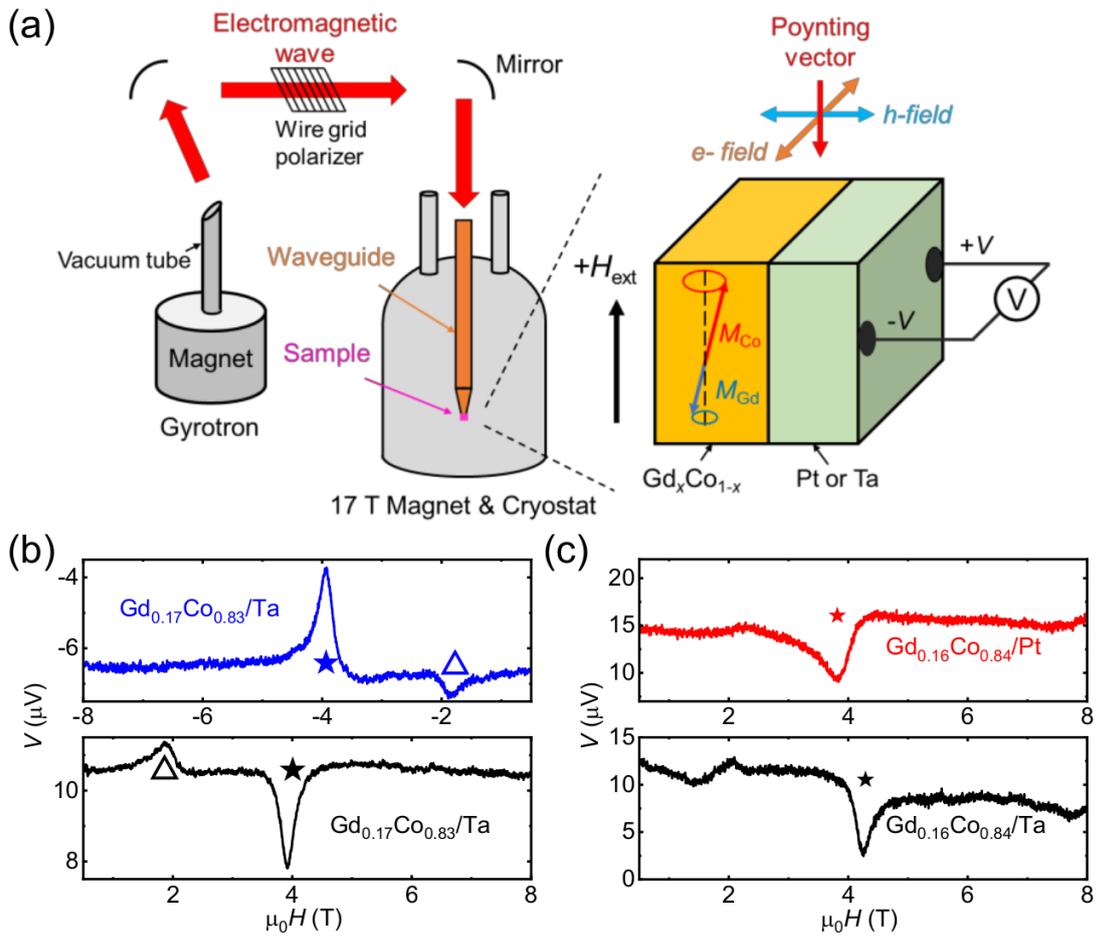

Fig. 1 Funada et al.

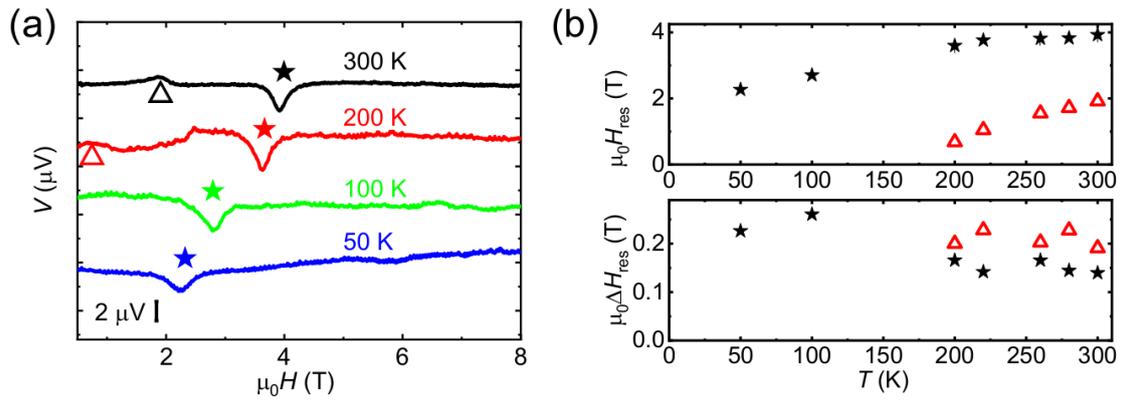

Fig. 2 Funada et al.

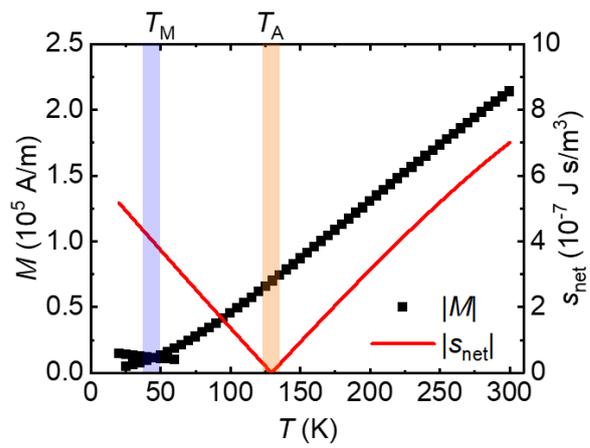

Fig. 3 Funada et al.

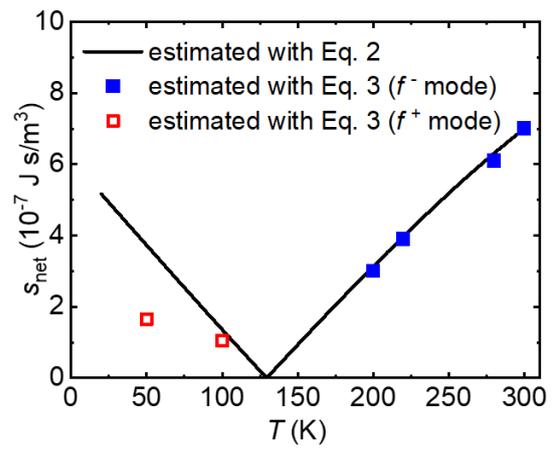

Fig. 4 Funada et al.

# References

[1] D. M. Mittleman, Perspective: Terahertz science and technology, J. Appl. Phys. **122**, 230901 (2017).

[2] V. Baltz, A. Manchon, M. Tsoi, T. Moriyama, T. Ono and Y. Tserkovnyak, Antiferromagnetic spintronics, Rev. Mod. Phys. **90**, 015005 (2018).

[3] T. Jungwirth, X. Marti, P. Wadley and J. Wunderlich, Antiferromagnetic spintronics, Nat. Nanotechnol. **11**, 231 (2016).

[4] O. Gomonay, T. Jungwirth and J. Sinova, Concepts of antiferromagnetic spintronics, Phys. Status Solidi Rapid Res. Lett. **11**, 1700022 (2017).

[5] S. K. Kim, G. S. D. Beach, K-J. Lee, T. Ono, T. Rasing, and H. Yang, Ferrimagnetic spintronics, Nat. Mater. **21**, 24 (2022).

[6] T. Kampfrath, A. Sell, G. Klatt, A. Pashkin, S. Ma, T. Dekorsy, M. Wolf, M. Fiebig, A. Leitenstorfer and R. Huber, Coherent terahertz control of antiferromagnetic spin waves, Nat. Photonics **5**, 31 (2011).

[7] T. Satoh, S. J. Cho, R. Iida, T. Shimura, K. Kuroda, H. Ueda, Y. Ueda, B. A. Ivanov, F. Nori and M. Fiebig, Spin Oscillations in Antiferromagnetic NiO Triggered by Circularly Polarized Light, Phys. Rev. Lett. **105**, 077402 (2010).

[8] S. Baierl, M. Hohenleutner, T. Kampfrath, A. K. Zvezdin, A. v. Kimel, R. Huber and R. v. Mikhaylovskiy, Nonlinear spin control by terahertz-driven anisotropy fields, Nat. Photonics **10**, 715 (2016).

[9] T. Moriyama, K. Hayashi, K. Yamada, M. Shima, Y. Ohya, Y. Tserkovnyak, and T. Ono, Enhanced antiferromagnetic resonance linewidth in NiO/Pt and NiO/Pd, Phys. Rev. B **101**, 060402(R) (2020).

[10] A. V. Kimel, B. A. Ivanov, R. V. Pisarev, P. A. Usachev, A. Kirilyuk and T. Rasing, Inertia-driven spin switching in antiferromagnets, Nat. Phys. 5, 727 (2009).

[11] S. Schlauderer, C. Lange, S. Baierl, T. Ebnet, C. P. Schmid, D. C. Valovcin, A. K. Zvezdin, A. V. Kimel, R.V. Mikhaylovskiy and R. Huber, Temporal and spectral fingerprints of ultrafast all-coherent spin switching, Nature **569**, 383 (2019).

[12] H. J. Juretschke, Electromagnetic Theory of dc Effects in Ferromagnetic Resonance, J. Appl. Phys. **31**, 1401 (1960).

[13] W. G. Egan and H. J. Juretschke, DC Detection of Ferromagnetic Resonance in Thin Nickel Films, J. Appl. Phys. **34**, 1477 (1963).

[14] N. Mecking, Y. S. Gui and C. M. Hu, Microwave photovoltage and photoresistance effects in ferromagnetic microstrips, Phys. Rev. B **76**, 224430 (2007).

[15] H. Mizuno, T. Moriyama, M. Kawaguchi, M. Nagata, K. Tanaka, T. Koyama, D. Chiba, and T. Ono, Ferromagnetic resonance measurements in sub-nanometer Fe films, Appl. Phys. Express 8, 073003 (2015).

[16] Luqiao Liu, Takahiro Moriyama, D. C. Ralph, and R. A. Buhrman, Spin-Torque Ferromagnetic Resonance Induced by the Spin Hall Effect, Phys. Rev. Lett. **106**, 036601 (2011).

[17] E. Saitoh, M.Ueda, and H. Miyajima, Conversion of spin current into charge current at room temperature: Inverse spin-Hall effect, Appl. Phys. Lett. 88, 182509 (2006)

[18] H. Nakayama, K. Ando, K. Harii, T. Yoshino, R. Takahashi, Y. Kajiwara, K. Uchida, Y. Fujikawa and E. Saitoh, Geometry dependence on inverse spin Hall effect induced by spin pumping in $Ni_{81}Fe_{19}$/Pt films, Phys. Rev. B **85**, 144408 (2012).

[19] J. Li, C. B. Wilson, R. Cheng, M. Lohmann, M. Kavand, W. Yuan, M. Aldosary, N. Agladze, P. Wei, M. S. Sherwin and J. Shi, Spin current from sub-terahertz-generated antiferromagnetic magnons, Nature **578**, 70 (2020).

[20] P. Vaidya, S. A. Morley, J. van Tol, Y. Liu, R. Cheng, A. Brataas, D. Lederman and E. del Barco, Subterahertz spin pumping from an insulating antiferromagnet, Science **368**, 160 (2020).

[21] T. Idehara and S. P. Sabchevski, Gyrotrons for High-Power Terahertz Science and Technology

at FIR UF, J. Infrared. Millim. Terahertz Waves **38**, 62 (2017).

[22] A. Fisher, Y. Park, M. Lenz, A. Ody, R. Agustsson, T. Hodgetts, A. Murokh and P. Musumeci, Single-pass high-efficiency terahertz free-electron laser, Nat. Photonics **16,** 441 (2022).

[23] C. D. Stanciu, A. v Kimel, F. Hansteen, A. Tsukamoto, A. Itoh and A. Kirilyuk, Ultrafast spin dynamics across compensation points in ferrimagnetic GdFeCo: The role of angular momentum compensation, Phys. Rev. B **73**, 220402(R) (2006).

[24] C. Kim, S. Lee, H. G. Kim, J. H. Park, K. W. Moon, J. Y. Park, J. M. Yuk, K. J. Lee, B. G. Park, S. K. Kim, K. J. Kim and C. Hwang, Distinct handedness of spin wave across the compensation temperatures of ferrimagnets, Nat. Mater. **19**, 980 (2020).

[25] K. J. Kim, S. K. Kim, Y. Hirata, S. H. Oh, T. Tono, D. H. Kim, T. Okuno, W. S. Ham, S. Kim, G. Go, Y. Tserkovnyak, A. Tsukamoto, T. Moriyama, K. J. Lee and T. Ono, Fast domain wall motion in the vicinity of the angular momentum compensation temperature of ferrimagnets, Nat. Mater. **16**, 1187 (2017).

[26] L. Caretta, M. Mann, F. Büttner, K. Ueda, B. Pfau, C. M. Günther, P. Hessing, A. Churikova, C. Klose, M. Schneider, D. Engel, C. Marcus, D. Bono, K. Bagschik, S. Eisebitt and G. S. D. Beach, Fast current-driven domain walls and small skyrmions in a compensated ferrimagnet, Nat. Nanotechnol. **13**, 1154 (2018).

[27] Y. Hirata, D. H. Kim, S. K. Kim, D. K. Lee, S. H. Oh, D. Y. Kim, T. Nishimura, T. Okuno, Y. Futakawa, H. Yoshikawa, A. Tsukamoto, Y. Tserkovnyak, Y. Shiota, T. Moriyama, S. B. Choe, K. J. Lee and T. Ono, Vanishing skyrmion Hall effect at the angular momentum compensation temperature of a ferrimagnet, Nat. Nanotechnol. **14**, 232 (2019).

[28] K. Ueda, M. Mann, P. W. P. de Brouwer, D. Bono, and G. S. D. Beach, Temperature dependence of spin-orbit torques across the magnetic compensation point in a ferrimagnetic TbCo alloy film, Phys. Rev. B **96**, 064410 (2017).

[29] J. Sinova, S. O. Valenzuela, J. Wunderlich, C. H. Back and T. Jungwirth, Spin Hall effects, Rev. Mod. Phys. **87**, 1213 (2015).

[30] A. Hoffmann, Spin Hall Effects in Metals, IEEE Trans Magn **49**, 5172 (2013).

[31] M. Harder, Z. X. Cao, Y. S. Gui, X. L. Fan and C. M. Hu, Analysis of the line shape of electrically detected ferromagnetic resonance, Phys. Rev. B **84**, 054223 (2011).

[32] T. A. Ostler, R. F. L. Evans, R. W. Chantrell, U. Atxitia, O. Chubykalo-Fesenko, I. Radu, R. Abrudan, F. Radu, A. Tsukamoto, A. Itoh, A. Kirilyuk, T. Rasing and A. Kimel, Crystallographically amorphous ferrimagnetic alloys: Comparing a localized atomistic spin model with experiments, Phys. Rev. B **84**, 024407 (2011).

[33] B. I. Min and Y. R. Jang, The effect of the spin-orbit interaction on the electronic structure of magnetic materials, J. Phys. Condens. **3**, 5131 (1991).

[34] F. Schlickeiser, U. Atxitia, S. Wienholdt, D. Hinzke, O. Chubykalo-Fesenko, and U. Nowak, Temperature dependence of the frequencies and effective damping parameters of ferrimagnetic resonance, Phys. Rev. B **86**, 214416 (2012).

[35] G. P. Rodrigue, H. Meyer and R. V. Jones, Resonance Measurements in Magnetic Garnets, J Appl Phys **31**, S376 (1960).

[36] S. K. Kim, K. Lee and Y. Tserkovnyak, Self-focusing skyrmion racetracks in ferrimagnets, Phys. Rev. B **95**, 140404(R) (2017).

[37] S. Funada, Y. Shiota, M. Ishibashi, T. Moriyama and T. Ono, Enhancement of spin wave group velocity in ferrimagnets with angular momentum compensation, Appl. Phys. Express **13**, 063003 (2020).

[38] S. K. Kim, Y. Tserkovnyak, and O. Tchernyshyov, Propulsion of a domain wall in an antiferromagnet by magnons, Phys. Rev. B **90**, 104406 (2014).

[39] See Supplemental Material at [URL will be inserted by publisher] for estimation of the magnetic susceptibility at $f_L$ and $f_R$.

[40] A. G. Gurevich and G. A. Melkov, *Magnetization oscillations and waves* (CRC Press, 1996).

[41] A. Kamra, R. E. Troncoso, W. Belzig and A. Brataas, Gilbert damping phenomenology for two-sublattice magnets, Phys. Rev. B **98**, 184402 (2018).

[42] J. Seib and M. Fähnle, Calculation of the Gilbert damping matrix at low scattering rates in Gd, Phys. Rev. B **82**, 064401 (2010).

[43] T. G. A. Verhagen, H. N. Tinkey, H. C. Overweg, M. van Son, M. Huber, J. M. van Ruitenbeek and J. Aarts, Temperature dependence of spin pumping and Gilbert damping in thin Co/Pt bilayers, J. Phys. Condens. **28**, 056004 (2016).

[44] N.H. Duc and D. Givord, Exchange interactions in amorphous Gd-Co alloys, J. Magn. Magn. Mater. **157**, 169 (1996).

[45] M. Huang, M. U. Hasan, K. Klyukin, D. Zhang, D. Lyu, P. Gargiani, M. Valvidares, S. Sheffels, A. Churikova, F. Büttner, J. Zehner, L. Caretta, K.-Y. Lee, J. Chang, J. -P. Wang, K. Leistner, B. Yildiz and G. S. D. Beach, Voltage control of ferrimagnetic order and voltage-assisted writing of ferrimagnetic spin textures, Nat. Nanotechnol. **16**, 981 (2021).